\documentclass[reprint,twocolumn,superscriptaddress,prx,longbibliography,groupedaddress]{revtex4-2}

\usepackage{amsmath,amssymb}
\usepackage{graphicx}% Include figure files
\usepackage{dcolumn}% Align table columns on decimal point
\usepackage{bm}% bold math
\usepackage{xr}
\usepackage[colorlinks = true,linktocpage=true]{hyperref}
\usepackage{tcolorbox}

\newcommand{\dt}{\mathrm{d}_t}

\begin{document}

\title{Flow of Energy and Information in Molecular Machines}

\author{Matthew P.\ Leighton}
\author{David A.\ Sivak}%
\affiliation{Department of Physics, Simon Fraser University, Burnaby, BC, V5A 1S6, Canada.}%
\email{matthew\_leighton@sfu.ca}
\email{dsivak@sfu.ca}

\date{\today}

%Abstract
\begin{abstract}
Molecular machines transduce free energy between different forms throughout all living organisms. While truly machines in their own right, unlike their macroscopic counterparts molecular machines are characterized by stochastic fluctuations, overdamped dynamics, and soft components, and operate far from thermodynamic equilibrium. In addition, information is a relevant free-energy resource for molecular machines, leading to new modes of operation for nanoscale engines. Towards the objective of engineering synthetic nanomachines, an important goal is to understand how molecular machines transduce free energy to perform their functions in biological systems. In this review we discuss the nonequilibrium thermodynamics of free-energy transduction within molecular machines, with a focus on quantifying energy and information flows between their components. We review results from theory, modeling, and inference from experiments that shed light on the internal thermodynamics of molecular machines, and ultimately explore what we can learn from considering these interactions.
\end{abstract}

\maketitle

%Table of Contents
\tableofcontents

\section{INTRODUCTION}
\label{sec:intro}
Living organisms operate far from thermodynamic equilibrium~\cite{schrodinger1992life}. Maintaining this nonequilibrium state requires a constant input of free energy, which must in turn be converted from one form to another in order to carry out the different functions necessary for life. At the cellular level, free-energy conversion is accomplished by \emph{molecular machines}: nanoscale protein structures responsible for much of the inner workings of the cell~\cite{Brown2019theory}.

\begin{figure*}[tb]
\includegraphics[width=\textwidth]{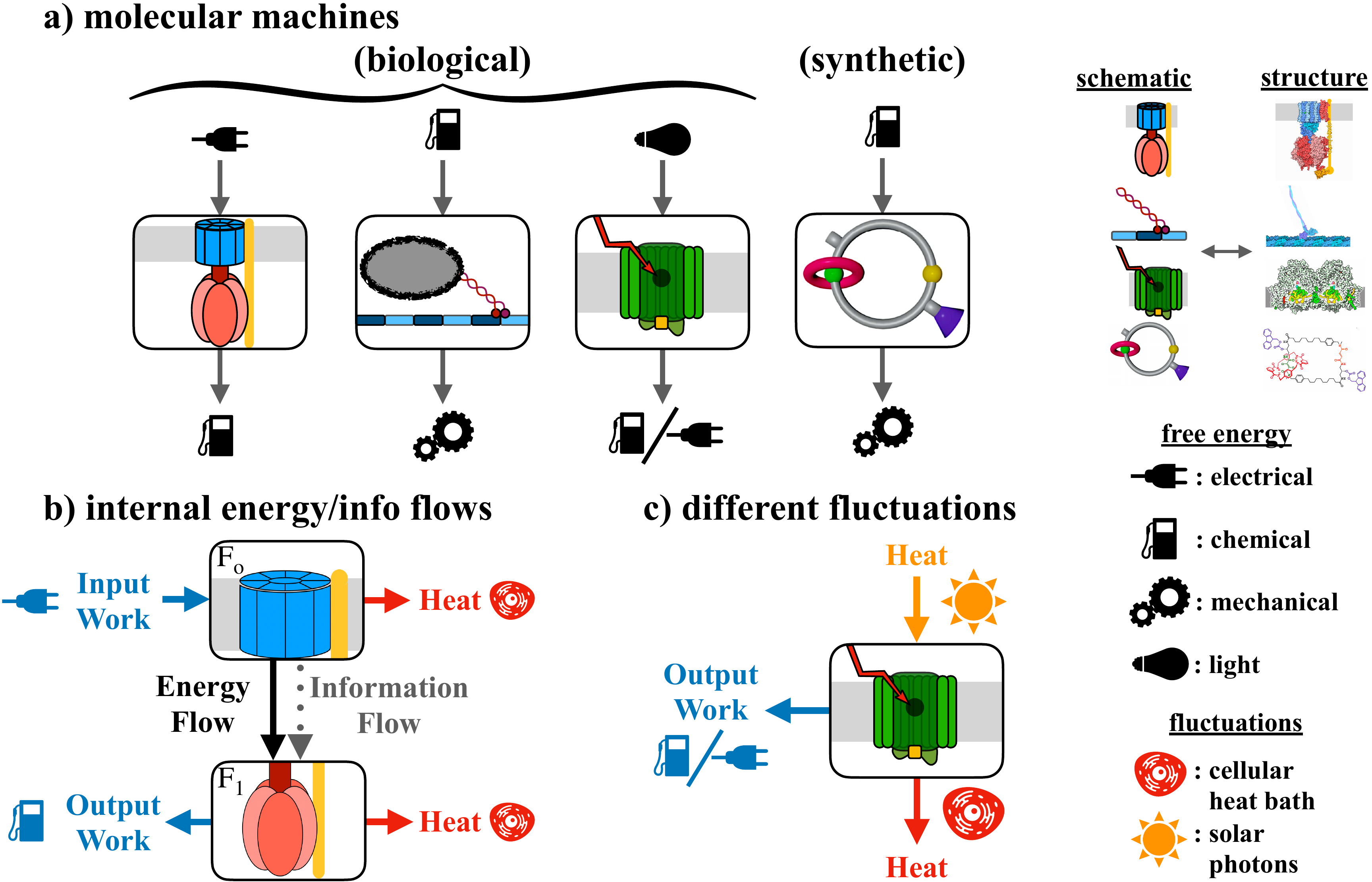}
\caption{Molecular machines transduce free energy between different forms at the nanoscale. a) Examples of biological and synthetic molecular machines and the types of free energy they transduce: from left to right, ATP synthase, kinesin, photosystem II (all structures adapted from the Protein Data Bank~\cite{berman2000protein}), and a minimal rotary motor (adapted from Ref.~\cite{amano2022insights_chemrxiv} (CC BY 4.0)). b) Modeling ATP synthase as bipartite, made up of the coupled rotary motors $\mathrm{F}_\mathrm{o}$ and $\mathrm{F}_1$, allows resolution of internal flows of energy and information. c) Some molecular machines, like photosystem II, are in contact with multiple sources of fluctuations, allowing them to effectively act as heat engines.}
\label{fig:fig1}
\end{figure*}

Biological molecular machines accomplish a wide variety of tasks within cells. Most living organisms on Earth ultimately derive their energy from the Sun, with high-energy solar photons being transduced into electrochemical free energy by photosynthetic organisms using light-harvesting molecular machines like photosystems I and II~\cite{vinyard2013photosystem}. Cells typically use molecules of adenosine triphosphate (ATP) as their energy currency, which is produced primarily by ATP synthases made of coupled rotary motors that leverage cross-membrane proton gradients to synthesize ATP against its chemical-potential gradient~\cite{oster1999atp}. ATP then powers a host of other molecular machines including transport motors like kinesin and myosin~\cite{howard2002mechanics}, protein synthesis machines like ribosomes, trans-membrane transporters such as sodium-potassium pumps, and a wide variety of machinery for DNA replication and transcription~\cite{phillips2012physical}. These molecules are truly machines, taking in energy and using it to perform useful work (Fig.~\ref{fig:fig1}a).

While they are indeed machines in their own right, molecular machines inhabit a very different physical world from the macroscopic machines we interact with in our everyday lives~\cite{Brown2019theory}. First of all, their dynamics are highly overdamped, with frictional drag dominating over inertia~\cite{purcell2014life}. Second, the energy scales of driving forces and interactions are comparable to the thermal energy scale $k_\mathrm{B}T$; thus their dynamics are strongly influenced by stochastic fluctuations from the environment. Third, molecular machines are made up of many soft protein components, coupled by loose, floppy connections. Finally, their strength of driving and rapidity of operation are such that they are at all times far from thermodynamic equilibrium.

Despite these bewildering differences, molecular machines perform remarkably well -- comparable to, and in some cases much better than, their macroscopic counterparts~\cite{marden2002molecules}. For example, photosystem II, one of the molecular machines responsible for photosynthesis in plant cells, converts solar energy into electrochemical free energy at an efficiency exceeding $40\%$~\cite{vinyard2013photosystem}. ATP synthase, a molecular machine inside the mitochondria, converts the electrical energy provided by a cross-membrane proton gradient into chemical free energy in the form of ATP, with an efficiency of $70$-$90\%$~\cite{silverstein2014exploration}. The $\mathrm{F}_\mathrm{o}$ and $\mathrm{F}_1$ subunits of ATP synthase are rotary motors that reach angular speeds of 150 rps~\cite{toyabe2010nonequilibrium}. Kinesin, a well-studied molecular transport motor, achieves velocities of up to $1\mu$m ($>100$ body lengths) per second~\cite{schnitzer1997kinesin}.

For nanoscale machines where energy scales are comparable to $k_{\rm B}T$, information (roughly speaking, correlated fluctuations) becomes a relevant thermodynamic resource. Molecular machines can interconvert between information and other types of free energy~\cite{Parrondo2015_Thermodynamics}. Conceptual models have been proposed for physically realizable molecular machines that could convert information directly into work~\cite{mcgrath2017biochemical}, and recent experiments have realized physical information-driven engines in laboratory conditions~\cite{serreli2007molecular,toyabe2010experimental,koski2014experimental,saha2021maximizing}. Such artificial \textit{information engines} can achieve velocities and power outputs comparable to biological molecular machines~\cite{saha2021maximizing}. While this information-transduction mechanism is understood and demonstrated in both theory and experiment, the question remains: have biological molecular machines evolved to use information as a thermodynamic resource?

In recent years, it has become possible to design and engineer synthetic molecular machines \textit{de novo}~\cite{balzani2000artificial,wilson2016autonomous,courbet2022computational,van2024coupled,korosec2024motility}; however, to date, these human-designed machines perform significantly worse than evolved biological molecular machines, for example achieving efficiencies of free-energy transduction on the order of $10^{-8}$~\cite{amano2022insights}. The practical promise of design principles to improve engineering of synthetic nanomachines drives the quest to better understand the inner workings of biological molecular machines. While the scientific community has learned much about biological molecular machines over the past decades, many questions remain about the physics underlying their function.

Like many synthetic molecular machines, biological molecular machines are typically made up of interacting components coupled together. Example include ATP synthase (composed of the coupled $\mathrm{F}_\mathrm{o}$ and $\mathrm{F}_1$ rotary motors~\cite{silverstein2014exploration}), the bacterial flagellar motor (a rotor coupled to $\sim$10 stators~\cite{meacci2009dynamics}), and assemblies of transport motors such as kinesin (where as many as hundreds of motors work collectively to pull large cargo~\cite{leopold1992association,rastogi2016maximum}). While single-molecule experiments and computational modeling have combined to yield an impressive understanding of how individual parts of molecular machines work in isolation~\cite{schnitzer1997kinesin,visscher1999single,toyabe2010nonequilibrium}, much less is known about how they interact with other components in their natural context as part of a larger conglomerate. Multicomponent molecular machines open new possibilities for free-energy transduction: in addition to exchanging free energy with external reservoirs, they also feature internal flows of free energy (comprised of energy and information) between their various coupled components~\cite{Ehrich2023_Energy}. Different components of multicomponent molecular machines can exchange energy with different sources of fluctuations (e.g., solar photons, active noise), leading in some cases to effective temperature differences across a machine~\cite{leighton2023information} (Fig.~\ref{fig:fig1}c).

\subsection{Stochastic Thermodynamics}
As free-energy conversion devices, it is natural and appealing to view molecular machines through the lens of thermodynamics; however, they present a significant challenge for application of classical thermodynamics as a result of their far-from-equilibrium operation, few degrees of freedom, and strongly stochastic dynamics. Motivated by these difficulties, the field of \emph{stochastic thermodynamics} has developed over the last 30 years to extend classical thermodynamics to nanoscale nonequilibrium stochastic systems. Stochastic thermodynamics provides a concrete framework for quantifying energy, heat, work, and other important quantities at the level of individual trajectories.

The first wave of important results included a collection of integrated and detailed fluctuation theorems, most notably those of Jarzynski~\cite{jarzynski1997nonequilibrium} and Crooks~\cite{crooks1999entropy}. These allowed, for the first time, direct estimation of equilibrium thermodynamic quantities like free-energy differences using data from experiments far from equilibrium~\cite{hummer2001free,liphardt2002equilibrium,collin2005verification}. The newfound tractability of nonequilibrium processes led to an explosion of work exploring different methods for obtaining optimal or near-optimal control protocols for probing nanoscale stochastic systems~\cite{schmiedl2007optimal,sivak2012thermodynamic,aurell2011optimal}.

More recently, interest has turned to bounding and inferring entropy production of nonequilibrium systems. Results such as the thermodynamic uncertainty relations~\cite{barato2015thermodynamic,gingrich2016dissipation,horowitz2020thermodynamic} and Jensen bounds~\cite{leighton2022dynamic,leighton2024jensen} provide lower bounds on entropy production rates of systems and subsystems in different contexts. In the study of molecular machines, these bounds have been used to obtain bounds and trade-offs constraining performance~\cite{barato2015thermodynamic,seifert2018stochastic,song2020thermodynamic,leighton2022dynamic,albaugh2023limits} and for thermodynamic inference of various efficiency measures~\cite{pietzonka2016universal,leighton2023inferring}.

Most importantly for the purpose of this review, the development of \emph{bipartite} stochastic thermodynamics has enabled a finer-resolution view of energy and entropy flows into, out of, and within more complex multicomponent stochastic systems~\cite{horowitz2014thermodynamics,barato2017thermodynamic}. This framework constitutes a natural lens through which to study energy and information flows in multicomponent molecular machines, and provides a firm grounding for the study of information thermodynamics in physically realizable systems.

%%%%%%%%%%%%%%%%%%%%%%%%%%%%%%%%%%%%%%%%%%%%
\subsection{Paradigmatic Molecular Machines}
Biology features a vast cornucopia of molecular machines. The  molecular machines that have been explored most thoroughly through the lens of nonequilibrium thermodynamics fall into several paradigmatic classes.

Some of the best-studied molecular motors are \emph{transport motors}, a large family of proteins that includes kinesins, dyneins, and myosins. These motors consume chemical free energy, generally in the form of ATP, which they transduce into mechanical energy by taking forward steps against mechanical forces~\cite{howard2002mechanics}. Another critically important class of molecular machines are \emph{rotary motors}, which feature rotational degrees of freedom driven by chemical, electrical, or mechanical forces. The family of rotary motors includes $\mathrm{F}_\mathrm{o}\mathrm{F}_1$ ATP synthase, responsible for the synthesis of ATP within mitochondria~\cite{oster1999atp}, V-ATPases, which serve as proton pumps across vacuolar membranes~\cite{beyenbach2006v}, and the bacterial flagellar motor~\cite{sowa2008bacterial}, the machinery underlying locomotion for many motile species of bacteria. Most biological free energy ultimately comes from sunlight, which is harvested by \emph{light-harvesting molecular machines} such as photosystems I and II in plant cells and bacteriorhodopsin in many archaea~\cite{phillips2012physical}. These molecular machines take in light energy in the form of solar photons, and transduce it into various different forms of electrochemical free energy. Biology also features a wide range of other machine types, including trans-membrane pumps, transporters, and the machinery involved in DNA and RNA transcription, translation, and repair~\cite{phillips2012physical}.

Beyond biological molecular machines, chemists and engineers have created a suite of synthetic molecular machines for various purposes~\cite{erbas2015artificial}. These are typically inspired by biological counterparts, but to date remain far simpler and much less functional. Work in this area has yielded synthetic versions of rotary motors~\cite{van2024coupled} and light-driven machines~\cite{corra2022kinetic,van2024coupled}, in addition to pumps, ratchets, and assemblers. Other lines of research involve computer simulations of models for synthetic machines, allowing rapid search of design space~\cite{amano2022insights,albaugh2022simulating}. Parallel efforts to improve our understanding of biological molecular machines strive to uncover design principles to guide these engineering efforts.

%%%%%%%%%%%%%%%%%%%%%%%%%%%%%%%%%%%%%%%%%%%%%
\subsection{Overview}
In this article, we review applications of stochastic thermodynamics to the study of molecular machines. We outline new theoretical developments that allow for quantifying energy and information flows into, out of, and within multi-component molecular machines, and review studies that have used this lens to examine molecular machines. Resolving internal flows of energy and information in molecular machines reveals a rich structure intimately connected to their external function and performance. We believe these tools will be indispensable, both for understanding the functioning of biological molecular machines and for learning design principles for engineering synthetic variants.

%%%%%%%%%%%%%%%%%%%%%%%%%%%%%%%%%%%%%%%%%%%%%
\subsection{Related Reviews}
This review focuses on quantifying flows of energy and information in molecular machines. Ref.~\cite{Brown2019theory} provides a broad theoretical perspective on free-energy transduction by molecular machines, while Ref.~\cite{li2020efficiencies} reviews measures of molecular-machine efficiency and Ref.~\cite{silverstein2014exploration} reviews the energetics of ATP synthase in different biological organisms. Ref.~\cite{mugnai2020theoretical} reviews theoretical considerations for modeling molecular motors. Ref.~\cite{sangchai2023artificial} reviews recent developments in engineering artificial molecular machines using information-ratchet mechanisms.

The theoretical aspects of this review are grounded in nonequilibrium thermodynamics, with review articles covering important theoretical frameworks including nonequilibrium work relations~\cite{Jarzynski2011_Equalities}, stochastic thermodynamics~\cite{seifert2012stochastic}, the thermodynamics of information~\cite{Parrondo2015_Thermodynamics}, optimal-control strategies~\cite{blaber2023optimal}, and a more pedagogical textbook introduction~\cite{Peliti2021_book}. For more recent developments in the field central to the content of this review, see reviews on thermodynamic inference~\cite{seifert2019stochastic}, thermodynamic uncertainty relations~\cite{horowitz2020thermodynamic}, bipartite stochastic thermodynamics~\cite{Ehrich2023_Energy}, and information flows~\cite{parrondo2023information}. From an experimental perspective, recent reviews summarize key experimental results in stochastic thermodynamics~\cite{ciliberto2017experiments} and experimental implementations of information engines~\cite{buisson2024performance}. Finally, Refs.~\cite{erbas2015artificial} and \cite{borsley2024molecular} review efforts to design synthetic molecular machines, while Refs.~\cite{hess2011engineering} and \cite{saper2019synthetic} review past and prospective engineering applications of biological molecular machines.

%%%%%%%%%%%%%%%%%%%%%%%
\section{NONEQUILIBRIUM THERMODYNAMICS OF STEADY STATES}
\label{sec:theory}
Classical thermodynamics is primarily concerned with heat engines operating cyclically in the quasistatic limit, or systems coming to equilibrium with external reservoirs; by contrast, molecular machines operate far from thermodynamic equilibrium. We are typically interested in their behavior in the long-time limit, which is dominated by the steady state. Thus, modeling molecular machines within the cellular environment requires a thermodynamics of nonequilibrium steady states.

As a paradigm for molecular machines, we consider a system in contact with an equilibrium thermal reservoir at temperature $T$ (and inverse temperature $\beta\equiv1/(k_\mathrm{B}T)$) with which it can exchange heat, as well as nonequilibrium free-energy reservoirs with which it can exchange work. Consider $\mathrm{F}_\mathrm{o}\mathrm{F}_1$-ATP synthase as a specific example, in which case the nonequilibrium free-energy reservoirs are the hydrogen-ion gradient across the mitochondrial membrane and nonequilibrium concentrations of ATP and ADP. 

As with any thermodynamic analysis, we start with first and second laws. While in classical thermodynamics we typically consider changes in energy and entropy along processes leading from one equilibrium state to another, here we consider steady-state rates of change for ensemble-averaged energy and entropy. Steady-state energy balance is quantified by the first law, $\dt{E} = \dot{W} + \dot{Q}$, with the rate of change of the internal energy $E$ equal to the sum of the rate of work $\dot{W}$ into the system from nonequilibrium free-energy reservoirs and the rate of heat $\dot{Q}$ into the system from the equilibrium thermal reservoir. Here $d_t$ denotes a total time derivative, while dots denote rates of path-dependent quantities like work and heat. The total entropy production rate due to the system dynamics is $\dot{\Sigma} = \dt{S} - \beta\dot{Q}\geq 0$, here given by the sum of the rate of change $\dt S$ of the system Shannon entropy and the rate $-\beta\dot{Q}$ of heat flow to the equilibrium thermal reservoir scaled by its temperature. For thermodynamically consistent discrete or continuous dynamics, the entropy production rate $\dot{\Sigma}$ is strictly nonnegative~\cite{seifert2012stochastic}.

\subsection{Bipartite Systems}
\label{sec:bipartite}
As discussed Sec.~\ref{sec:intro}, many molecular machines of interest are composed of multiple coupled components. Thermodynamically, different components of molecular machines can be treated as coupled subsystems, each of which are thermodynamic systems in their own right. The simplest example consists of a system $Z$ composed of two subsystems $X$ and $Y$. These subsystems are \emph{bipartite} if their degrees of freedom can evolve independently, and are subject to independent fluctuations from their environments. This means that each subsystem can be treated as being in exclusive contact with a distinct thermal reservoir. More precisely, this condition requires (for discrete degrees of freedom) that $X$ and $Y$ cannot simultaneously change states, or (for continuous degrees of freedom) that the noise terms in the Langevin equations describing the temporal evolution of $X$ and $Y$ are uncorrelated~\cite{Ehrich2023_Energy}.

In a bipartite system, flows of energy and entropy are resolvable at the level of individual subsystems. The $X$ and $Y$ subsystems exchange work $\dot{W}_{X/Y}$ and heat $\dot{Q}_{X/Y}$ with their respective reservoirs, and exchange energy as what we call the \emph{energy flow}:
\begin{equation}
\dot{E}_{Y} \equiv \lim_{\tau\to0}\frac{E\left[X(t),Y(t+\tau)\right] - E\left[X(t),Y(t)\right]}{\tau}.
\end{equation}
This is the rate at which the dynamics of $Y$ increase the shared energy $E[X,Y]$ ($\dot{E}_X$ is defined analogously). In this sense $\dot{E}_Y$ is an \emph{energy flow} from $Y$ to $X$. At steady state the energy is constant, so any increase in energy due to $Y$ dynamics must be accompanied by a matching decrease in energy due to $X$:
$\dot{E}_X = -\dot{E}_Y$. These energy flows are related by \emph{local first laws} describing energy balance for each subsystem at steady state:
\begin{subequations}\label{eq:steadystatefirstlaws}
\begin{align}
\dot{E}_Y & = \dot{W}_Y + \dot{Q}_Y,\\
\dot{E}_X & = \dot{W}_X + \dot{Q}_X.
\end{align}
\end{subequations}
Analogously, the subsystem-level entropy flows are resolvable by decomposing the total system entropy production rate into nonnegative contributions from each subsystem~\cite{Ehrich2023_Energy}:
\begin{subequations}\label{eq:steadystatesecondlaws}
\begin{align}
\dot{\Sigma}_Y & = - \beta_Y\dot{Q}_Y - \dot{I}_Y\geq 0,\\
\dot{\Sigma}_X & = - \beta_X\dot{Q}_X - \dot{I}_X\geq 0.
\end{align}
\end{subequations}
Here the entropy production rate $\dot{\Sigma}_X$ ($\dot{\Sigma}_Y$) of the $X$ ($Y$) subsystem is the sum of a contribution $\beta_X\dot{Q}_X$ ($\beta_Y\dot{Q}_Y$) due to heat flow to its associated thermal reservoir, and the \emph{information flow} $\dot{I}_Y$. The information flow (alternatively known as the learning rate~\cite{brittain2017we}) is the rate of change of the mutual information $I[X,Y]$~\cite{Cover2006_Elements} between $X$ and $Y$, due to the dynamics of $Y$, formally defined as
\begin{equation}\label{eq:infoflowdefs}
\dot{I}_Y \equiv \lim_{\tau\to0}\frac{I\left[X(t),Y(t+\tau)\right] - I\left[X(t),Y(t)\right]}{\tau}.
\end{equation}
The information flow $\dot{I}_X$ is defined analogously, and---as with the energy flow---at steady state the mutual information is constant, so $\dot{I}_X = -\dot{I}_Y$. 

\begin{tcolorbox}
\textbf{MAKING SENSE OF THE INFORMATION FLOW}

Compared to the energy flow $\dot{E}_Y$, the information flow $\dot{I}_Y$ is a more nebulous thermodynamic quantity. The information flow is a component of the total change in system free energy due to the dynamics of $Y$, which we call the \emph{transduced free energy}. At steady state $Y$ increases the free energy $F = E - k_\mathrm{B}TS$ of the system at a rate
\begin{subequations}
\begin{align}
\dot{F}_Y & \equiv \dot{E}_Y - k_\mathrm{B}T\dot{S}_Y\\
& = \dot{E}_Y + k_\mathrm{B}T\dot{I}_Y.
\end{align}
\end{subequations}
$\dot{S}_Y$ is the rate at which $Y$ increases the system entropy. At steady state the marginal entropy $S[Y] = S[X,Y] + I[X,Y] - S[X]$ is constant, so that the entropic part of the transduced free energy (quantifying changes in joint entropy) is given by the information flow~\cite{Ehrich2023_Energy}. 

The information flow is the rate at which the dynamics of the $Y$ subsystem increase the mutual information $I[X,Y]$ between the two subsystems. This mutual information is a thermodynamic resource, which can be used by, e.g., the $X$ subsystem: if the $X$ dynamics decrease the mutual information ($\dot{I}_X<0$), then the subsystem second law allows $X$ to entirely convert heat from its environment ($\dot{Q}_X>0$) into output work ($-\dot{W}_X>0$). If only $X$ is observed, it appears to convert heat directly into work, seemingly a violation of the second law. Only when the information flow is taken into account, does the validity of the second law become apparent. Thus a useful way of thinking about the information flow $\dot{I}_Y$ is as the capacity of subsystem $X$ to convert heat into work.
\end{tcolorbox}

Figure~\ref{fig:fig1}b illustrates an example of the bipartite setup, showing the energy and information flows for the $\mathrm{F}_\mathrm{o}$ and $\mathrm{F}_1$ subsystems of $\mathrm{F}_\mathrm{o}\mathrm{F}_1$-ATP synthase. Together, the energy and information flow (multiplied by the temperature) constitute the free energy flowing from one subsystem to the other, with the information flow constituting the entropic component.

\begin{figure*}[tb]
\includegraphics[width=\textwidth]{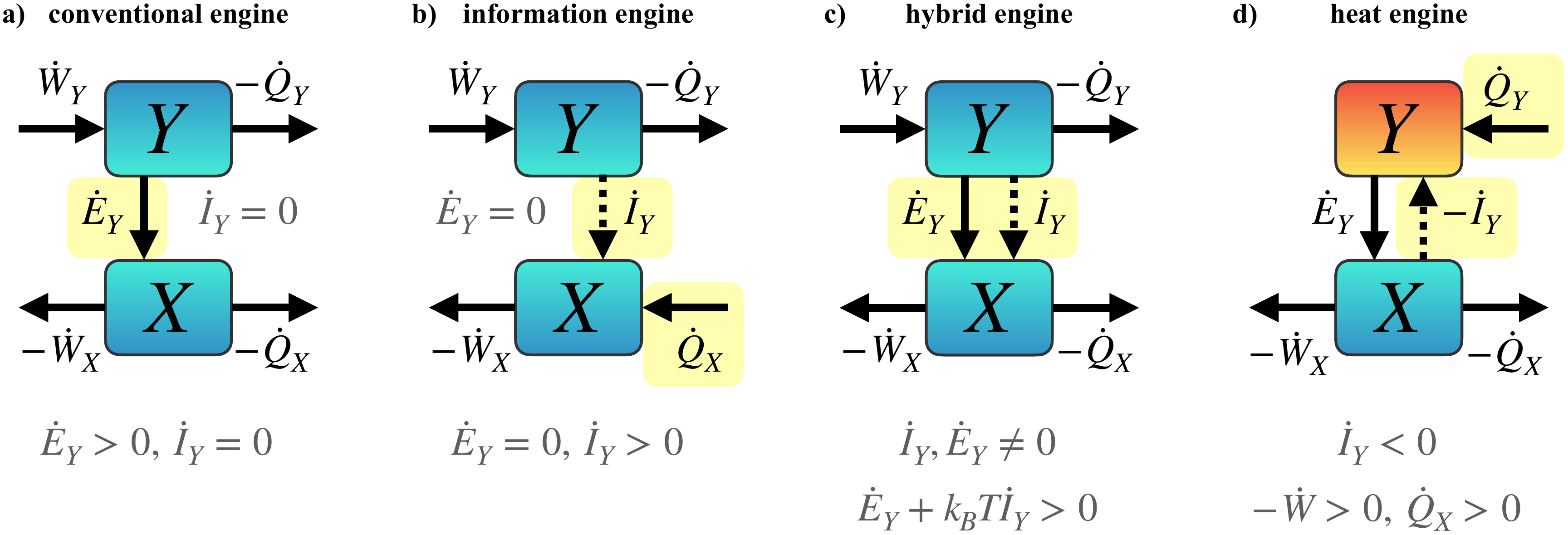}
\caption{Different operational modes of bipartite thermodynamic engines, with their defining constraints below. a) Conventional engine. b) Information engine. c) Hybrid engine. d) Heat engine ($T_Y>T_X$). Arrows show direction of energy and information flows, all symbols are positive quantities for the operational modes depicted.}
\label{fig:fig2}
\end{figure*}

These thermodynamic laws constrain the possible flows within bipartite engines. Figure~\ref{fig:fig2} illustrates several example operational modes that generate output work $-\dot{W}_X$. One such example is the \textit{conventional engine}, where free energy from input work $\dot{W}_Y$ is transduced solely in the form of energy flow between the $Y$ and $X$ subsystems. In this case, one subsystem takes in input work, the other produces output work, and both subsystems dissipate heat to their environments. Alternatively, it is possible to construct a pure \textit{information engine}, where free energy is instead exchanged solely via information flow: one subsystem (here $Y$) takes input work, creates mutual information between $Y$ and $X$, and dissipates heat $-\dot{Q}_Y$ to the environment; the other subsystem then uses the mutual information to extract heat from the environment, which it then turns into output work $-\dot{W}_X$.

Conventional and information engines represent extremes, each utilizing only one type of free-energy transduction. More generally, work converters feature both internal energy and information flows (Fig.~\ref{fig:fig2}c); we call these \textit{hybrid} engines. Finally, bipartite machines with access to different sources of fluctuations can operate as \textit{heat engines} (Fig.~\ref{fig:fig2}d), producing output power by leveraging a temperature difference. As discussed in Sec.~\ref{sec:infoflows}, the directionality of internal information and energy flows is constrained~\cite{leighton2023information}.

\subsection{Efficiency Metrics for Bipartite Machines}
\subsubsection{Work Transducers}
Many molecular machines transduce free energy from one form to another, taking in input work $\dot{W}_Y$ via one subsystem ($Y$, without loss of generality), and outputting work $-\dot{W}_X$ via the other subsystem. Examples include the conventional, information, and hybrid engines in Fig.~\ref{fig:fig2}. For work transducers, thermodynamic efficiency can be defined in a straightforward manner as the ratio of output to input work,
\begin{equation}\label{eq:thermoeff}
\eta_\mathrm{T} \equiv \frac{-\dot{W}_X}{\dot{W}_Y} \leq 1.
\end{equation}

The bipartite structure allows us to resolve more fine-grained details of free-energy transduction \emph{within} the system. Combining the first (Eq.~\eqref{eq:steadystatefirstlaws}) and second (Eq.~\eqref{eq:steadystatesecondlaws}) laws of bipartite stochastic thermodynamics at steady state yields two nested inequalities relating the free-energy inputs and outputs of each subsystem~\cite{barato2017thermodynamic,lathouwers2022internal}:
\begin{equation}
\dot{W}_Y \geq \dot{E}_Y + k_{\rm B}T \dot{I}_Y\geq -\dot{W}_X.
\end{equation}
Here the sum of the energy and information flows, the internally transduced free energy $\dot{E}_Y + k_{\rm B}T \dot{I}_Y$, acts as a bottleneck between input and output work. This motivates the introduction of subsystem efficiencies~\cite{barato2017thermodynamic,leighton2023information}, which quantify the efficiency of free-energy transduction for each subsystem:
\begin{subequations}
\begin{align}
\eta_Y & \equiv \frac{\dot{E}_Y + k_{\rm B}T \dot{I}_Y}{\dot{W}_Y} \leq 1,\label{eq:ysubsystemeff}\\
\eta_X & \equiv \frac{-\dot{W}_X}{\dot{E}_Y + k_{\rm B}T \dot{I}_Y} \leq 1.
\end{align}
\end{subequations}
Their product is the thermodynamic efficiency of the whole system, $\eta_\mathrm{T} = \eta_Y\eta_X$. Other subsystem efficiency definitions have also been proposed to quantify the efficiency of converting information into heat energy~\cite{horowitz2014thermodynamics}.

\subsubsection{Transport Motors}
Many molecular machines, rather than transducing energy from one form to another, consume free energy to transport cargo from one place to another. Examples include transport motors such as kinesin pulling a diffusive cargo, and the flagellar motor pushing a cell. For these kinds of systems, we can think of the motor and cargo as the $Y$ and $X$ subsystems, respectively. In this case, the output of the motor is motion of the cargo at some average velocity $\langle v_X\rangle$ against the force of viscous friction, for Stokes flow equal to the friction coefficient $\zeta_X$ times the velocity. The average work rate against this viscous friction force is thus $\zeta_X\langle v_X\rangle^2$, suggesting the definition of the \emph{Stokes efficiency}~\cite{wang2002stokes}:
\begin{equation}\label{eq:stokeseff}
\eta_\mathrm{S} \equiv \frac{\zeta_X\left\langle v_X\right\rangle^2}{\dot{W}_Y} \leq 1.
\end{equation}
For transport systems, the motor itself still has a well-defined subsystem efficiency (Eq.~\eqref{eq:ysubsystemeff}), quantifying how much of the input work is made available to the cargo as transduced free energy~\cite{leighton2023inferring}.

\subsection{Generalizations}
\subsubsection{Multipartite Systems}\label{multipartite}
While Sec.~\ref{sec:bipartite} focuses on bipartite systems, many of the main results generalize to multipartite systems with $N>2$ degrees of freedom. As with the bipartite assumption, a system is multipartite when all subsystems are influenced by independent fluctuations, with the same mathematical consequences as in the bipartite case. Many molecular machines of interest are well-described as multipartite systems, for example collective transport systems where anywhere from one to over a hundred transport motors jointly pull a single cargo~\cite{leighton2022performance,leighton2022dynamic}, or the bacterial flagellar motor where $\approx 10$ stators collectively apply torque to turn a single rotor.

In a multipartite system, each subsystem satisfies a first law, $\dot{E}_i = \dot{W}_i + \dot{Q}_i$, and a second law~\cite{horowitz2015multipartite}, $\dot{\Sigma}_i = - \beta_i\dot{Q}_i - \dot{I}_i$. Here the energy flow $\dot{E}_i$ and information flow $\dot{I}_i$ are respectively the rates of change of the system internal energy and the total correlation~\cite{crooks2017measures} due to the dynamics of the $i$th subsystem.

While the framework detailed above is best suited to the study of bipartite or multipartite stochastic systems, efforts have been made to extend certain results to systems lacking the bipartite structure~\cite{chetrite2019information,leighton2023information}.

\subsubsection{Chemical Reaction Networks}
While our focus in this review is on analysis of single molecular machines, when large collections are considered together they can be modeled using deterministic chemical reaction networks, with state concentrations replacing state probabilities as the relevant variables. Nonequilibrium steady-state thermodynamics can still be formulated for chemical reaction networks~\cite{rao2016nonequilibrium}, and when these networks are bipartite the energy and information flows can be defined as for single molecular machines~\cite{penocchio2022information}.

%%%%%%%%%%%%%%%%%%%%%%%
\section{MODELING FREE-ENERGY TRANSDUCTION IN MOLECULAR MACHINES}
\label{sec:modelling}

\subsection{Rotary Motors}
Rotary motors with a single degree of freedom have been workhorse model systems in the development of stochastic thermodynamics. Studying their dynamics and thermodynamics has led to general principles~\cite{tu2018design,kasper2020modeling}, and improved our understanding of specific biological motors like $\mathrm{F}_1$~\cite{toyabe2010nonequilibrium,toyabe2011thermodynamic,kawaguchi2014nonequilibrium,gupta2022optimal} and $\mathrm{F}_\mathrm{o}$~\cite{zhai2024power}. Biological rotary motors, such as ATP synthase and the bacterial flagellar motor, are composite systems generally made up of multiple rotary components coupled together. To this end, recent work has begun to investigate the dynamics and thermodynamics of coupled rotary motors. While a wide range of minimal models have been considered~\cite{xing2005continuum,golubeva2012efficiency,fogedby2017minimal}, here we focus our attention on those concretely related to biological molecular machines.

\begin{figure*}[tb]
\includegraphics[width=\textwidth]{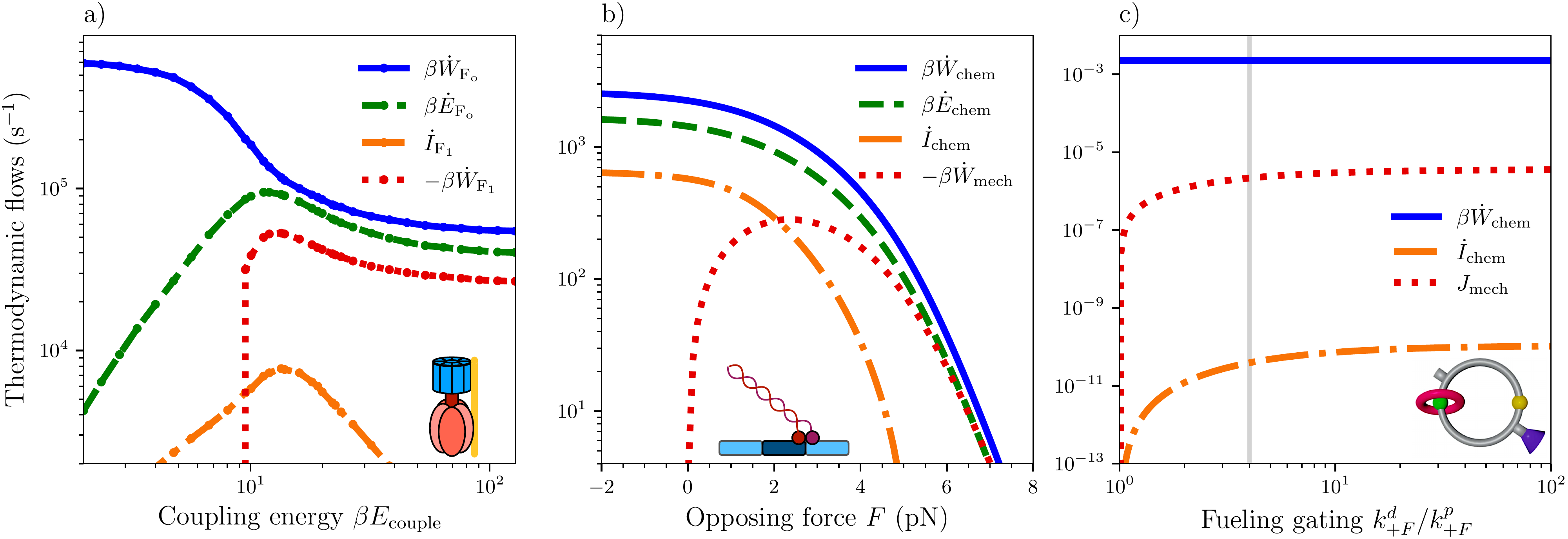}
\caption{Insights from analysis of the energy and information flows in simple models of molecular machines. a) Energy and information flows for $\mathrm{F}_\mathrm{o}$ and $\mathrm{F}_1$ in ATP synthase, as functions of the coupling energy $\beta E_\mathrm{couple}$, from the model considered in Ref.~\cite{lathouwers2022internal}. b) Energy and information flows from the chemical and mechanical components of the kinesin motor model studied in Ref.~\cite{Takaki2022_Information}, as functions of the opposing force $F$. c) Chemical-work consumption, information flow, and mechanical flux in a model, as functions of the fueling gating, for the synthetic rotary motor studied in Ref.~\cite{amano2022insights_chemrxiv}. Gray vertical line indicates experimental conditions. Panel a) adapted with permission from Ref.~\cite{lathouwers2022internal}; copyrighted by the American Physical Society. Inset schematic and code for c) adapted from Ref.~\cite{amano2022insights_chemrxiv} (CC BY 4.0).}
\label{fig:fig3}
\end{figure*}

The coupled rotary motors of ATP synthase constitute a textbook example of a bipartite molecular machine. The rotational degrees of freedom in ATP synthase, the c-ring in $\mathrm{F}_\mathrm{o}$ and the $\gamma$-shaft in $\mathrm{F}_1$, can be modeled using overdamped Langevin equations coupled by a rotationally symmetric joint potential energy~\cite{okazaki2015elasticity,lathouwers2020nonequilibrium,lathouwers2022internal}. One of the most notable findings~\cite{lathouwers2020nonequilibrium} is that output power depends non-monotonically on the coupling strength between $\mathrm{F}_\mathrm{o}$ and $\mathrm{F}_1$. While efficiency is highest for tight coupling, the output power is maximized at intermediate coupling strength (Fig.~\ref{fig:fig3}a). This somewhat counterintuitive result can be understood by considering the dominant transition paths (inchworming and slippage) during coupled rotation of the two motors~\cite{lathouwers2020nonequilibrium}: at low coupling strength the motors are decoupled, while at high coupling strength they can only rotate in lockstep; intermediate coupling strength allows enough flexibility for $\mathrm{F}_\mathrm{o}$ and $\mathrm{F}_1$ to separately cross their respective energy barriers (inchworming) while keeping slippage minimal.  

Exploring a more detailed picture of energy and information flows sheds further light on the inner workings of ATP synthase. Figure~\ref{fig:fig3}a shows the input and output work rates $\dot{W}_{\mathrm{F}_\mathrm{o}}$ and $-\dot{W}_{\mathrm{F}_1}$, along with the energy and information flows $\dot{E}_{\mathrm{F}_\mathrm{o}}$ and $\dot{I}_{\mathrm{F}_1}$. As with the output power, the transduced free energy and energy flow from $\mathrm{F}_\mathrm{o}$ to $\mathrm{F}_1$ are both maximized at intermediate coupling, along with the information flow. This information flow from $\mathrm{F}_1$ to $\mathrm{F}_\mathrm{o}$ allows $\mathrm{F}_\mathrm{o}$ to decrease the heat it dissipates to the environment, thereby increasing the energy transduced to $\mathrm{F}_1$~\cite{lathouwers2022internal}.

Closely related to $\mathrm{F}_\mathrm{o}\mathrm{F}_1$-ATP synthase, $\mathrm{V}_\mathrm{o}\mathrm{V}_1$-ATPases are similarly composed of two coupled rotary motors and have recently been the subject of experimental investigation~\cite{kosugi2023design}, but have yet to be studied through the lens of stochastic thermodynamics. Another prominent example is the bacterial flagellar motor, which drives flagellar rotation and enables locomotion in many species of motile bacteria. Similar to ATP synthase, the bacterial flagellar motor is made up of multiple coupled rotary motors: a central rotor, surrounded by a varying number of stators. While the dynamics of this motor complex have been studied through stochastic models~\cite{meacci2009dynamics,cao2022modeling}, much less is known from a thermodynamic standpoint. In particular, we see significant promise in exploring free-energy transduction between the different components of the bacterial flagellar motor. 

\subsection{Transport Motors}
The stochastic thermodynamics community 
has also devoted significant attention to the collection of motor proteins (e.g., kinesin, dynein, and myosin) used to transport materials within cells. The prototypical system is a single kinesin motor pulling a diffusive cargo (e.g., a vesicle) subject to an applied external force. More generally, transport systems can include anywhere from 1-10~\cite{shtridelman2008force,shtridelman2009vivo} to well over 100~\cite{leopold1992association} motors of different types working collectively to transport large cargos. Transport motor proteins are also used for other purposes, for example applying forces on actin filaments within muscle tissue~\cite{rastogi2016maximum}.

These systems feature many different types of energy and information flows. At the smallest scale, Ref.~\cite{Takaki2022_Information} considered flows of energy and information between the heads of individual kinesin and myosin motors, as well as changes in energy and information due to contributions from chemical and mechanical dynamics. Moving up in scale, energy and information flows between motor(s) and cargo can also be considered, for example for single motors pulling externally controlled cargo~\cite{ariga2018nonequilibrium}, collections of motors pulling diffusive cargo~\cite{leighton2022performance,leighton2022dynamic}, and collections of motors working against external loads~\cite{wagoner2021evolution}.

Figure~\ref{fig:fig3}b illustrates the thermodynamics of the discrete model for kinesin stepping considered in Ref.~\cite{Takaki2022_Information}. Shown are the input chemical work rate $\dot{W}_\mathrm{chem}$, the output mechanical work rate $-\dot{W}_\mathrm{mech}$, and the changes due to the chemical dynamics in the internal energy, $\dot{E}_\mathrm{chem}$, and the mutual information (of the chemical states of the motor's two heads), $\dot{I}_\mathrm{chem}$. This model shows significant information flow for small opposing forces, with information constituting an important part of the transduced free energy when output power is maximized. This internal information flow is interpreted as a quantitative measure of ``gating", an allosteric interaction mediated by mechanical strain between motor heads where the chemical state of one head regulates the stepping kinetics of the other head~\cite{yildiz2008intramolecular}. Gating is believed to increase both processivity~\cite{andreasson2015examining} and the stall force~\cite{hinczewski2013design}.

While the thermodynamics of free-energy transduction within motors, and between motors and their cargo, are now fairly well-explored, thus far less attention has been given to the energy and information flows between motors working either cooperatively or in opposition. Recent modeling studies and experiments predict cooperative binding interactions for both myosin motors on actin filaments~\cite{wagoner2021evolution} and kinesin motors on microtubules~\cite{wijeratne2022motor}. We believe it would be of significant interest to explore these interactions through the lens of information flows between motors. This direction is especially intriguing in light of recent work on engineering collective-transport systems with precisely controlled motor counts~\cite{delrosso2017exploiting,derr2012tug}.

\subsection{Synthetic Molecular Machines}
With the engineering of synthetic molecular machines becoming increasingly possible, it is natural to ask whether this framework of quantifying free-energy transduction between components of molecular machines can prove useful as a tool for improving the design of synthetic molecular machines. Here we highlight studies of synthetic molecular machines through the lens of stochastic thermodynamics.

The first nanoscale autonomous chemically fueled molecular motor~\cite{wilson2016autonomous} was analyzed from a thermodynamic standpoint in Ref.~\cite{amano2022insights}. This rotary motor (Fig.~\ref{fig:fig1}a fourth panel) consists of two interlocking molecular rings with two ``docking stations" where motion of the smaller ring is alternately blocked or permitted by binding/unbinding of chemical fuel. The motor was modeled as bipartite, with coupled chemical and mechanical degrees of freedom. In this case, the experimentally synthesized molecular motor was shown to operate as an information ratchet, with free energy transduced entirely via information flow $\dot{I}_\mathrm{chem}$ (Fig.~\ref{fig:fig3}c). While this motor is strikingly less efficient than biological molecular machines, with the chemical subsystem's efficiency on the order of $10^{-8}$, an exploration in parameter space around the physically realized model allowed the authors to determine design principles for future improvements. For example, by adjusting the ``fueling gating" (the ratio of two rate constants, quantifying attachment bias of the chemical fuel), the mechanical flux $J_\mathrm{mech}$ can be increased at constant chemical input work $\dot{W}_\mathrm{chem}$ (Fig.~\ref{fig:fig3}c). This increase in performance occurs via a corresponding increase in the information flow, and thus the subsystem efficiency (Eq.~\eqref{eq:ysubsystemeff}) of free-energy transduction from the chemical to mechanical subsystems.

This framework has since been employed to study other synthetic molecular motors. Analysis of a chemically driven information ratchet used the lens of internal energy and information flows to explore the role of kinetic asymmetry and power strokes as design parameters for adjusting machine performance~\cite{binks2023role}, showing that these parameters can tune both the magnitude of free-energy transduction as well as the relative roles of energy and information flows.  Bipartite thermodynamic analysis of a light-driven supramolecular pump quantified entropy production in different machine components~\cite{corra2022kinetic}. We expect that analyzing energy and information flows will prove useful for engineering other molecular machines.

%%%%%%%%%%%%%%%%%%%%%%%
\section{COMPARATIVE ADVANTAGE OF INFORMATION FLOW}
\label{sec:infoflows}
As demonstrated in Sec.~\ref{sec:modelling}, information flows can generically be found in simple models for molecular machines. The models of ATP synthase and kinesin studied in Refs.~\cite{lathouwers2022internal} and \cite{Takaki2022_Information} respectively show significant, or indeed even maximal, information flows in parameter regimes where the machines produce maximum output power. Likewise, simulations of models for synthetic molecular machines also find significant information flow~\cite{amano2022insights,binks2023role}. In other cases however, for example a discrete-state model for ATP synthase~\cite{Grelier2023_Unlocking}, information flow is exactly zero when output power is maximized. The family of continuous transport motor models considered in Refs.~\cite{leighton2022performance} and \cite{leighton2022dynamic} likewise yield optimal performance (saturating the Jensen entropy-production bound, Sec.~\ref{sec:Jensen}) for potential-energy landscapes which result in no information flows between motors and cargo. It remains an open question whether, and if so under what general conditions, information flows provide an advantage. 

As discussed in Sec.~\ref{sec:theory}, information flows between subsystems allow for parts of a system to seemingly break the second law of thermodynamics by converting heat from the environment into useful output work. In the most extreme case, information flow allows a machine to act as an information engine, where free energy is transduced from one part of the system to the other entirely via information. This allows machine components to be free-energetically coupled without energetic coupling. This is leveraged in synthetic molecular machines such as the one developed in Ref.~\cite{wilson2016autonomous}, where even despite a lack of energetic \emph{coupling}, researchers achieved biased motion via information flow alone~\cite{amano2022insights}.

\subsection{Leveraging Different Sources of Fluctuations}
While the question of when information flows should be used remains open in full generality, recent work has shed light on one context in which information flows provide an advantage: when molecular machines contact different sources of fluctuations. The cellular environment is typically assumed to be isothermal, with identical fluctuations acting on every part of any given molecular machine. While this is indeed the case for many systems of interest, there are also many cases in which fluctuations may not be completely homogeneous. 

One such example is light-harvesting molecular machines like photosystem II and bacteriorhodopsin, which on one hand interact with thermal and chemical fluctuations at the ambient cellular temperature, and on the other are driven by high-energy fluctuations from solar radiation. The monochromatic radiation with which these molecular machines interact can be treated as high-temperature thermal fluctuations, specifically at the temperature whose blackbody spectrum produces the intensity of that radiation~\cite{buddhiraju2018thermodynamic,penocchio2021nonequilibrium,corra2022kinetic}. Another example coming into focus is that of active fluctuations powered by metabolic activity, which are now appreciated to permeate the cellular interior~\cite{Mizuno2007_Nonequilibrium,Gallet2009_Power,pietzonka2019autonomous,dabelow2019irreversibility}. These fluctuations can differ significantly from basic thermal fluctuations in both their intensity in certain frequencies and their correlation structure. Additionally, even molecular machines like ATP synthase which operate across membranes may have access to small temperature gradients on the order of $1-10$K~\cite{Pinol2020_Real-Time,macherel2021conundrum,Wu2022_Intracellular,Di2022_Spatiotemporally}. 

The existence of these different sources of fluctuations raises the question of whether biological molecular machines can leverage them to improve performance. A molecular machine with components driven by different sources of fluctuations can be modeled as a \emph{bipartite heat engine}, with the different subsystems in contact with thermal reservoirs at different temperatures. These temperatures quantify the relationship between heat flow, information flow, and entropy production in the subsystem second laws (Eq.~\eqref{eq:steadystatesecondlaws}). 

While for isothermal work transducers the transduced free energy $\dot{E}_Y + k_{\rm B}T\dot{I}_Y$ is a bottleneck constraining both the input and output work rates, for a bipartite heat engine the work rates are instead constrained by two separate inequalities~\cite{Grelier2023_Unlocking}:
\begin{subequations}
\begin{align}
\dot{W}_Y  \geq & \dot{E}_Y + k_{\rm B}T_Y\dot{I}_Y, \\
 & \dot{E}_Y + k_{\rm B}T_X\dot{I}_Y \geq -\dot{W}_X.
\end{align}
\end{subequations}
Here the temperature weights the free-energetic contribution of the information flow relative to the energy flow. As a result of the different temperatures,  energy and information have different relative values for the two subsystems. This can be understood intuitively through analogy to economic arbitrage~\cite{leighton2023information}. 

Combining these two inequalities yields the \emph{information flow arbitrage relation} (IFAR)~\cite{leighton2023information}:
\begin{equation}\label{eq:ifar}
-\dot{W}_Y -\dot{W}_X \leq k_{\rm B}(T_X-T_Y)\dot{I}_Y.
\end{equation}
Here the net output work rate is upper bounded by a product of the temperature difference and the information flow $\dot{I}_Y$. Thus a bipartite heat engine must use information flow to take advantage of a temperature difference to generate net output work; moreover, the information must flow from the cooler subsystem to the hotter one.

The conclusion that information flow should be used to optimize performance under a temperature difference is further reinforced by Ref.~\cite{Grelier2023_Unlocking}, which explored adding a temperature difference to a discrete-state model for coupled rotary motors inspired by ATP synthase. Under isothermal conditions, output power is maximized for parameters that result in zero information flow, but with a temperature difference, however, maximizing output power produces substantial information flow. As predicted by the IFAR, the directionality of information flow is from the colder subsystem to the hotter one.

\subsection{Leveraging Active Fluctuations}
While active fluctuations can in some cases be modeled as thermal fluctuations at a higher temperature, this is certainly not always the case. Active fluctuations can have significant spatiotemporal correlations, which is likely the case for biologically relevant active noise~\cite{Mizuno2007_Nonequilibrium}. 

A general theory has yet to be developed for thermodynamic engines leveraging active fluctuations; nevertheless, explorations of specific systems have led to insights. Outside of molecular machines, recent work has shown that information can be used to leverage active fluctuations to produce useful work in active heat engines~\cite{datta2022second}, active matter~\cite{vansaders2023informational}, and microorganisms swimming through turbulent flows~\cite{monthiller2022surfing}. Of particular interest, Ref.~\cite{datta2022second} derives a second-law-like inequality for active engines which, similar to the IFAR, bounds output work in terms of information-theoretic flows.

For molecular machines, active fluctuations as a resource remain relatively unexplored. Intriguing experiments and computation have shown that certain types of active noise can speed up kinesin operation when pulling a diffusive cargo against high load forces~\cite{ariga2021noise}. We hope future work will explore both the underlying thermodynamics of this behavior, and whether other molecular machines can similarly leverage active noise to improve their performance.

%%%%%%%%%%%%%%%%%%%%%%%
\section{THERMODYNAMIC INFERENCE}
\label{sec:inference}
Many thermodynamic quantities of interest, for example efficiencies and internal energy and information flows, are difficult to quantify experimentally. Calculating these quantities generally requires full knowledge of the joint nonequilibrium probability distribution governing the relevant degrees of freedom, along with knowledge of all conservative and nonconservative forces acting on the system. Such information is typically beyond our current ability to measure for systems as complex as biological molecular machines, and can be difficult to compute even for numerical simulations of model systems. As such, we turn to thermodynamic inference: the study, development, and application of mathematical tools to infer hidden thermodynamic properties from quantities that can be observed experimentally~\cite{seifert2019stochastic}. In this section we outline a collection of recently developed theoretical tools for inference that have been applied to study various molecular machines.

\subsection{Entropy-Production Bounds and Their Applications}
Bounds on entropy production due to the dynamics of a system are particularly useful tools for thermodynamic inference. One generic example is bounding the steady-state thermodynamic efficiency of an isothermal molecular machine converting between different forms of work. Typical single-molecule experiments probe motor proteins working against applied load forces, and measure the output work rate $\dot{W}_\mathrm{out}$ but not the input work rate $\dot{W}_\mathrm{in}$. In this case, a lower bound $\dot{\Sigma}_\mathrm{LB}$ on the entropy production rate leads to an upper bound on the thermodynamic efficiency, $\eta_\mathrm{T} \leq 1/\left(1 + \dot{\Sigma}_\mathrm{LB}/\dot{W}_\mathrm{out}\right)$.
At a more fine-grained level, lower bounds on subsystem entropy production rates can be used, in combination with knowledge of the two work rates, to derive pairs of lower and upper bounds on subsystem efficiencies in bipartite molecular machines~\cite{leighton2023inferring}:
\begin{subequations}
\begin{align}\label{eq:subeffbounds}
\eta_\mathrm{T}\left(1 + \frac{\dot{\Sigma}^\mathrm{LB}_X}{-\beta\dot{W}_X}\right) & \leq \eta_Y \leq 1 - \frac{\dot{\Sigma}^\mathrm{LB}_Y}{\beta\dot{W}_Y},\\
\eta_\mathrm{T}\left(1 - \frac{\dot{\Sigma}^\mathrm{LB}_Y}{\beta\dot{W}_Y}\right)^{-1} & \leq \eta_X \leq \left(1 + \frac{\dot{\Sigma}^\mathrm{LB}_X}{-\beta\dot{W}_X}\right)^{-1}.
\end{align}
\end{subequations}
Note that in both of these examples, upper bounds on entropy production could instead be employed to obtain different formulations of efficiency bounds~\cite{leighton2023inferring}. An important goal in stochastic thermodynamics is then to derive entropy production bounds, calculable from available data, which can be inserted into relationships such as these.

\subsubsection{Thermodynamic Uncertainty Relations}
A widely used entropy production bound emerging from stochastic thermodynamics is the family of thermodynamic uncertainty relations (TURs)~\cite{horowitz2020thermodynamic}. These inequalities relate entropy production to the means and variances of time-integrated dynamic or thermodynamic currents. TURs typically take the form
\begin{equation}\label{eq:tur}
\Sigma_\tau \geq \frac{2\left\langle J_\tau\right\rangle}{\mathrm{Var}\left(J_\tau\right)},
\end{equation}
where $\Sigma_\tau$ is the entropy production over a time interval $\tau$, and $J_\tau$ is any suitably defined current. In the short-time limit, bounds on the entropy production rate are obtained.

\begin{figure*}[tb]
\includegraphics[width=\textwidth]{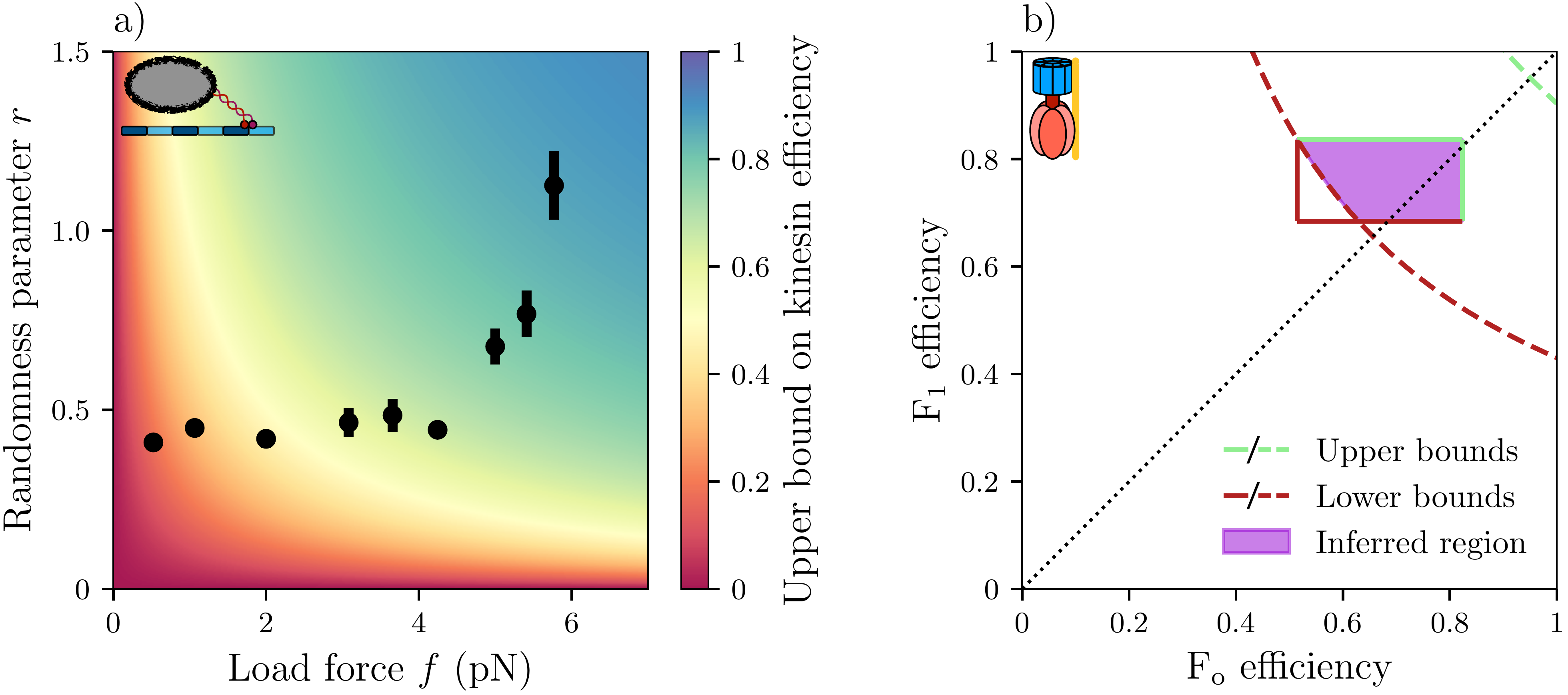}
\caption{Thermodynamic inference of the efficiencies of molecular machines. a) Inferring the efficiency of a kinesin motor pulling against a variable applied force using the TUR. Black points denote experimental data from Ref.~\cite{visscher1999single}. Color quantifies upper bound on the thermodynamic efficiency of the motor.
Panel adapted with permission from Ref.~\cite{seifert2018stochastic}.
b) Inferred subsystem efficiencies of $\mathrm{F}_\mathrm{o}$ and $\mathrm{F}_1$ in ATP synthase using the subsystem Jensen bounds.
Panel adapted with permission from Ref.~\cite{leighton2023inferring}; copyrighted by the American Physical Society.}
\label{fig:fig4fig}
\end{figure*}

An immediate application of the TUR, first demonstrated in Ref.~\cite{pietzonka2016universal}, is inferring the efficiencies of molecular transport motors from measurements of their average velocity $\left\langle v\right\rangle$ and effective diffusivity 
\begin{equation}
D_\mathrm{eff}\equiv\lim_{t\to\infty}\frac{\left\langle\Delta x^2\right\rangle - \left\langle \Delta x\right\rangle^2}{2t}.
\end{equation}
In particular, the thermodynamic efficiency (Eq.~\eqref{eq:thermoeff}) of a motor pulling against a constant load force $f$, and the Stokes efficiency (Eq.~\eqref{eq:stokeseff}) of a motor pulling a diffusive cargo against viscous drag, are upper bounded by
\begin{subequations}
\begin{align}
\eta_\mathrm{T} & \leq \left(1 + \frac{\left\langle v\right\rangle}{\beta D_\mathrm{eff} f_\mathrm{load}}\right)^{-1},\label{eq:TURetabound}\\
\eta_\mathrm{S} & \leq \frac{D_\mathrm{eff}}{D_\mathrm{c}}.
\end{align}
\end{subequations}
Figure~\ref{fig:fig4fig}a illustrates an application of Eq.~\eqref{eq:TURetabound}, showing maximum kinesin efficiency for different values of the load force $f$ and randomness parameter $r\equiv 2D_\mathrm{eff}/\left(\left\langle v\right\rangle d\right)$ ($d$ is the step distance of the motor). Plotting experimental kinesin data from Ref.~\cite{visscher1999single} shows that the maximum kinesin efficiency compatible with the data increases with the load force~\cite{seifert2018stochastic}. This inference is performed without requiring any knowledge of the motor's chemical dynamics.

The TUR can also be used to quantitatively assess the performance of molecular machines, by measuring how close their true entropy production rate comes to saturating the bound. This has been done, for example, for a simulated artificial molecular motor, revealing significant deviation from the bound~\cite{albaugh2023limits}. Investigating the source of these inefficiencies can uncover design principles for improving performance.

Beyond the canonical form of the TUR (Eq.~\eqref{eq:tur}), other related bounds include TUR-type bounds on subsystem entropy production rates valid at short times~\cite{otsubo2020estimating}, and a pair of bipartite TUR bounds for subsystem entropy production rates that hold more generally~\cite{tanogami2023universal}. We see significant potential for application of these results to thermodynamic inference in molecular machines.

\subsubsection{Jensen Bounds}
\label{sec:Jensen}
Another family of entropy production bounds are the Jensen bounds~\cite{leighton2022dynamic,leighton2024jensen}. For stochastic systems with continuous degrees of freedom $\{x_i\}_{i=1}^N$ obeying multipartite overdamped or underdamped Langevin dynamics with constant diffusion coefficients $D_i$, this framework yields lower bounds on both subsystem and total entropy production rates~\cite{leighton2024jensen}:
\begin{subequations}
\begin{align}
\dot{\Sigma}_i & \geq \frac{\left\langle \dot{x}_i\right\rangle^2}{D_i},\label{eq:localjensen}\\
\dot{\Sigma} = \sum_{i}\dot{\Sigma}_i & \geq \sum_{i=1}^N \frac{\left\langle \dot{x}_i\right\rangle^2}{D_i}.\label{eq:globaljensen}
\end{align}
\end{subequations}
These bounds are derived by applying Jensen's inequality to expressions for subsystem entropy production rates in multipartite systems~\cite{Horowitz2015_Multipartite,dechant2018entropic}, but in some cases can also be derived via a short-time version of the TUR~\cite{otsubo2020estimating}. These Jensen bounds can also be generalized to allow for position-dependent diffusion coefficients and non-multipartite dynamics~\cite{leighton2024jensen}.

The subsystem entropy production bounds (Eq.~\eqref{eq:localjensen}) are particularly well-suited to quantitative inference of free-energy transduction within molecular machines. Inserted into the subsystem efficiency bounds (Eq.~\eqref{eq:subeffbounds}), using the example of $\mathrm{F}_\mathrm{o}\mathrm{F}_1$ ATP synthase which tightly couples rotation and electrochemical free-energy consumption, one obtains the following bounds constraining the efficiencies of the $\mathrm{F}_\mathrm{o}$ and $\mathrm{F}_1$ rotary motors~\cite{leighton2023inferring}:
\begin{subequations}
\begin{align}
\eta_\mathrm{T}\left(1-\frac{\zeta_1\left\langle J_1\right\rangle}{\Delta\mu_\mathrm{ATP}}\right) & \leq \eta_\mathrm{o}\leq 1-\frac{\zeta_\mathrm{o}\left\langle J_\mathrm{o}\right\rangle}{\Delta\mu_{\mathrm{H}^+}},\\
\eta_\mathrm{T}\left(1-\frac{\zeta_\mathrm{o}\left\langle J_\mathrm{o}\right\rangle}{\Delta\mu_{\mathrm{H}^+}}\right)^{-1} & \leq \eta_1 \leq \left(1-\frac{\zeta_1\left\langle J_1\right\rangle}{\Delta\mu_\mathrm{ATP}}\right)^{-1}.
\end{align}
\end{subequations}
These bounds allow for quantitative inference of $\eta_\mathrm{o}$ and $\eta_1$ from experimentally accessible quantities: friction coefficients $\zeta_{\mathrm{o}/1}$, average rotation rates $\langle J_{\mathrm{o}/1}\rangle$, and chemical driving forces $\Delta\mu_{\mathrm{H}^+/\mathrm{ATP}}$. Inserting estimates of the various parameters in the above bounds yields Fig.~\ref{fig:fig4fig}b, showing that this method can tightly constrain the subsystem efficiencies of $\mathrm{F}_\mathrm{o}$ and $\mathrm{F}_1$. In contrast to single-molecule experiments measuring the energetics of a single motor (e.g., $\mathrm{F}_1$) in isolation~\cite{toyabe2010nonequilibrium,toyabe2011thermodynamic}, these bounds allow inference of free-energy transduction between $\mathrm{F}_\mathrm{o}$ and $\mathrm{F}_1$ in their natural coupled context. This framework also allows inference of subsystem efficiencies in other bipartite systems, for example a kinesin motor pulling a diffusive cargo~\cite{leighton2023inferring}.

Like the TUR, the Jensen bounds can also be employed to quantify performance of molecular machines. Ref.~\cite{leighton2022dynamic} applied the Jensen bound to derive Pareto frontiers constraining combinations of performance metrics like velocity, efficiency, and power consumption in collective transport systems. For example, a collection of $N$ identical motors pulling a diffusive cargo satisfies a velocity-efficiency inequality, $\eta + \left\langle v\right\rangle/v_\mathrm{max}\leq 1$. Here $v_\mathrm{max}$ is the maximum velocity of a single unloaded motor, and $\eta$ can be either of the thermodynamic or Stokes efficiencies. This inequality can also be inverted to infer efficiency from measurements of $\left\langle v\right\rangle$ and $v_\mathrm{max}$.

\subsubsection{Other Bounds on Entropy Production}
While the TURs and Jensen bounds have seen applications towards inference for molecular machines, many other bounds on entropy production have been derived through the formalism of stochastic thermodynamics. Most prominently, these include a collection of \emph{thermodynamic speed limits}~\cite{shiraishi2018speed,falasco2020dissipation,van2020unified,nakazato2021geometrical} that lower bound the entropy production in terms of the time taken to move a system from one probability distribution to another. 

While constraints on entropy production typically provide lower bounds, Dechant~\cite{dechant2023upper} recently derived an upper bound on the entropy production rate of a stochastic system,
$\dot{\Sigma} \leq \beta \left\langle\bm{F}_\mathrm{nc}^\top \bm{\mu}\bm{F}_\mathrm{nc}\right\rangle$,
in terms of an ensemble-averaged norm of the non-conservative forces $\bm{F}_\mathrm{nc}$.
Intuitively, the magnitude of nonconservative forces limits how far from equilibrium the system can be driven, thus limiting the entropy production rate. Combined with previously derived lower bounds on entropy production, upper bounds hold promise for more precise inference.

Entropy production can also be inferred directly, for example using the \emph{Variance Sum Rule} (VSR)~\cite{di2024variance,di2024variance2}, which requires measurements of the second time derivative of the variance of position displacement along with the variance of the forces acting on the system. While the VSR has yet to be applied to the study of molecular machines, we believe it has significant potential as an inference tool.

\subsection{Arbitrage Relations for Inferring Energy and Information Flows}
\begin{figure}[h]
\includegraphics[width=\columnwidth]{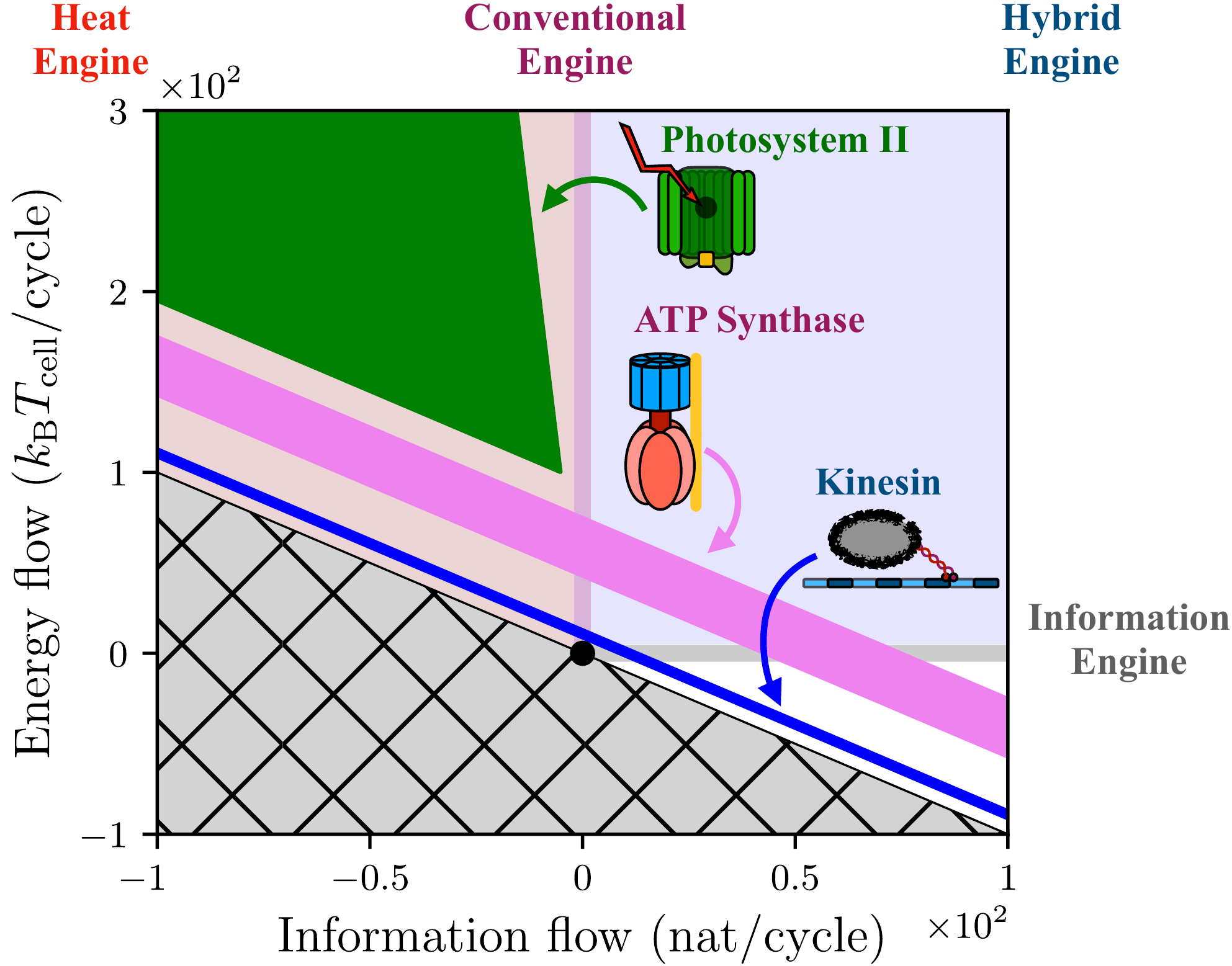}
\caption{Physical constraints on internal energy flows (relative to the thermal energy at cellular temperature $T_\mathrm{cell}$) and information flows (in units of nats~\cite{Cover2006_Elements}) in molecular machines. Gray hatched region indicates negative free energy transduction $\dot{E}_Y + k_{\rm B}T_\mathrm{cell}\dot{I}_Y<0$. Black circle: equilibrium ($\dot{E}_Y = 0 = \dot{I}_Y$). Light shaded regions show operational modes of heat, conventional, hybrid, and information engines (clockwise). Darker shaded regions show constraints on energy and information flows in the molecular machines photosystem II, ATP synthase, and kinesin.}
\label{fig:fig5}
\end{figure}

The arbitrage relations~\cite{leighton2023information} discussed in Sec.~\ref{sec:infoflows} can also be used for thermodynamic inference, implying the existence and directionality of energy and information flows in bipartite systems operating as heat engines. When $T_X>T_Y$, the information flow is bounded by
\begin{equation}\label{eq:infoarbitrageinference}
\dot{I}_Y \geq \frac{-\dot{W}}{k_{\rm B}(T_X-T_Y)}.
\end{equation}
Analogously, the energy flow is bounded by
\begin{equation}
\dot{E}_X \geq \frac{-\dot{W}_X/T_X - \dot{W}_Y/T_Y}{1/T_Y - 1/T_X}.
\end{equation}
Equation~\eqref{eq:infoarbitrageinference} has been applied to infer information flows in the light-harvesting molecular machines photosystem II and bacteriorhodhopsin, yielding respective quantitative estimates of $\sim7$ and $\sim0.5$ bits per reaction cycle~\cite{leighton2023information}, consistent with direct calculations from numerical simulations of experimentally parameterized models. Figure~\ref{fig:fig5} shows the constraints from arbitrage relations on the energy and information flows in photosystem II. 

\subsection{Comparing Free-Energy Transduction Strategies in Distinct Machines}
Combining published results on thermodynamic inference allows us to sketch out a map of free-energy transduction strategies used by different biological molecular machines. Figure~\ref{fig:fig5} shows inferred energy and information flows in ATP synthase, kinesin, and photosystem II using results from Refs.~\cite{leighton2023inferring} and \cite{leighton2023information}. The $Y$ and $X$ subsystems correspond respectively to $\mathrm{F}_\mathrm{o}$ and $\mathrm{F}_1$ (ATP synthase), kinesin and its diffusive cargo (kinesin), and the P680 and OEC subunits (photosystem II), assigned such that $X$ is in contact with the cellular heat bath at $T_\mathrm{cell}$ and free energy flows from $Y$ to $X$, hence $\dot{E}_Y + k_\mathrm{B}T_\mathrm{cell}\dot{I}_Y\geq 0$.

We show regions for heat, conventional, hybrid, and information engines, along with inferred constraints on flows inside select molecular machines. For photosystem II, the arbitrage relations constrain its internal thermodynamics to the heat-engine regime, while for ATP synthase and kinesin, constraints from the subsystem Jensen bounds leave open all four different operational modes as possibilities. We hope future work will add further constraints and other molecular machines to this figure. Figure~\ref{fig:fig5} also clearly illustrates how these machines differ in the magnitude of internal free-energy transduction (quantitatively, distance from the bounding line where $\dot{E}_Y + k_\mathrm{B}T_\mathrm{cell}\dot{I}_Y= 0$): photosystem II supports significantly more internal free-energy transduction per cycle than ATP synthase, which in turn out-transduces kinesin by a similar margin.

%%%%%%%%%%%%%%%%%%%%%%%
\section{CONCLUSION}
\label{sec:discussion}
In this review, we explored the thermodynamics of free-energy transduction by molecular machines, with a special focus on internal flows of energy and information between their components. Traditional analyses consider free-energy flows into and out of a machine. A finer-grained view, decomposing the machine into multiple coupled components, resolves internal details -- internal flows of energy and information that enable the machine's function. 

Bipartite stochastic thermodynamics~\cite{Ehrich2023_Energy} permits quantification of energy and information flows in two-component stochastic systems, and imposes first (Eq.~\eqref{eq:steadystatefirstlaws}) and second (Eq.~\eqref{eq:steadystatesecondlaws}) laws on each component of a molecular machine. These describe energy and entropy balance at the level of individual subsystems, and lead to meaningful constraints on the possible operational modes of multicomponent stochastic thermodynamic engines (see Fig.~\ref{fig:fig2}).

In Sec.~\ref{sec:modelling}, we highlighted several examples of bipartite models for molecular machines including ATP synthase, kinesin, and a synthetic rotary motor (see Fig.~\ref{fig:fig3}). The lens of internal energy and information flows leads to new kinds of analysis in these systems. For example, in ATP synthase, output power is maximized for an intermediate strength of the coupling between $\mathrm{F}_\mathrm{o}$ and $\mathrm{F}_1$, which in turn maximizes transduced free energy and allows for significant information flow~\cite{lathouwers2022internal}. In synthetic molecular machines, the bipartite lens helps in exploring design space, showing that tuning parameters to increase information flow leads to increased mechanical flux~\cite{amano2022insights}.

The information flow, which arises from proper accounting of entropy balance in composite stochastic systems, is a feature of particular interest: somewhat mysterious, but essential for a proper thermodynamic understanding of nanoscale machines~\cite{parrondo2023information}. Reviewing work on various model systems shows a wide range of qualitatively different information flows in different contexts, with a shortage of general principles for when information provides advantages. For molecular machines with access to different sources of fluctuations, recent theoretical work (Sec.~\ref{sec:infoflows}) has shown that information flows are required to generate net output work.

\begin{tcolorbox}\textbf{SUMMARY POINTS}

\begin{enumerate}
\item Molecular machines transduce free energy from one external reservoir to another via internal energy and information flows between coupled components. 
\item Energy and information flows can be calculated directly from numerical simulations of models for molecular machines, allowing for exploration of how they relate to overall performance.
\item Information flows allow molecular machines to 
take advantage of different sources of fluctuations in a manner reminiscent of heat engines.
\item Hidden thermodynamic quantities like internal energy and information flows as well as various measures of efficiency can be inferred indirectly from experimental measurements using results from stochastic thermodynamics.
\end{enumerate}
\end{tcolorbox}

While internal energy and information flows and other thermodynamic quantities (including various measures of efficiency) can be computed from simulations of stochastic models for molecular machines, directly measuring these quantities in experiment is much more difficult, and frequently intractable for current techniques: experimental determination of such quantities currently requires thermodynamic inference~\cite{seifert2019stochastic}. Section~\ref{sec:inference} illustrated techniques for inferring thermodynamic quantities using inequalities from stochastic thermodynamics like entropy production bounds and arbitrage relations, or equalities such as the variance sum rule. We highlight particular results including bounds on efficiencies of kinesin, $\mathrm{F}_\mathrm{o}$, and $\mathrm{F}_1$ (Fig.~\ref{fig:fig4fig}), and bounds on the information flow in photosystem II (Fig.~\ref{fig:fig5}). Such results start to constrain possible operational modes used by biological molecular machines.

\subsection{Future Directions}
Investigations using simple models for biological molecular machines like ATP synthase and kinesin have begun to give us glimpses into their inner workings. Future work should develop more sophisticated models, taking advantage of increasingly precise and plentiful experimental data to more closely match reality. Increasingly accurate models instill greater confidence that their behavior matches the hidden details of the machines they correspond to, and that operational modes (see Fig.~\ref{fig:fig5}) have been inferred correctly. Another important next step is to develop models to tackle other classes of molecular machines that have yet to be considered through the lens of stochastic thermodynamics, for example membrane transporters. Exploring free-energy transduction, and especially information flows, in models for the various machines involved in the biological central dogma of replication, transcription, and translation would be of significant interest; indeed, preliminary work suggests that RNA polymerase may operate as a Maxwell demon~\cite{tsuruyama2023rna}. It would also be interesting to explore applications to allostery in catalytic enzymes, where interactions between spatially separated subunits are central to their functionality, and connections to molecular machines have been drawn~\cite{fourmond2021reversible}. 

Another important area for both theoretical and modeling work is the information thermodynamics of transient behavior, beyond the nonequilibrium steady states considered in this review. This regime is relevant for molecular machines subject to short-timescale environmental changes, or external control. As detailed in Sec.~\ref{sec:infoflows}, developing general theory for molecular machines subject to highly correlated active noise is also an area demanding further exploration.

The paradigmatic molecular machines discussed throughout this review can all be plausibly modeled as bipartite. Many other molecular machines of significant biophysical interest, however, do not invite such a simple decomposition. For example, the bacterial flagellar motor is composed of $\mathcal{O}(10)$ coupled components, while the ribosome is thought to comprise $\mathcal{O}(100)$. As such, future theoretical work should seek to solidify and expand the framework of multipartite stochastic thermodynamics beyond the first and second laws described in Sec.~\ref{multipartite}, for example towards multipartite generalizations of subsystem efficiencies. While for bipartite systems the internal energy and information flows can be described as flowing from one subsystem to another, in a multipartite system such an interpretation is much less straightforward to make except in special cases where the subsystems are sparsely connected~\cite{leighton2022performance}. Going from a single system to a bipartite one introduces the possibility of internal energy and information flows; it is natural to wonder what new qualitative features arise when systems are decomposed further. We expect the topology of the graph describing interactions between components to play an important role.

The information flow discussed throughout this review is at this point fairly well-understood mathematically, but still somewhat mysterious from the standpoint of physical intuition. For example, Ref.~\cite{leighton2023information} inferred $\gtrsim7$ bits information flow per reaction cycle in photosystem II. How should we interpret the magnitude of this information flow? The information flow can be interpreted in terms of measurement and feedback~\cite{Ehrich2022_Energetic}, but the connection to an autonomous molecular machine could be sharpened. Future work should develop intuition for the magnitude of information flows, perhaps drawing on connections between stochastic thermodynamics and computation~\cite{wolpert2023stochastic}.

In parallel to advancing modeling efforts, future work should develop more and better tools for thermodynamic inference. While work to date, outlined in Sec.~\ref{sec:inference}, has yielded constraints on the internal details of specific biological machines, these results are insufficient to determine the operational modes used in isothermal machines like kinesin and ATP synthase. Tighter bounds and alternative methods are needed to improve both the precision and accuracy of estimates. Likewise, future work should bring these inference tools to bear on other molecular machines beyond those that have been studied thus far. Exploring the internal details of different classes of molecular machines with improved inference tools will more systematically identify the free-energy transduction strategies used by different classes of evolved biological molecular machines (see Fig.~\ref{fig:fig5}) in different contexts, allowing identification of patterns and extraction of design principles for engineering synthetic molecular machines.

For experimentalists, near-term future work should strive to measure quantities that can be used for thermodynamic inference (Sec.~\ref{sec:inference}). These include friction coefficients and rates of change for continuous degrees of freedom (e.g., angular coordinates in rotary motors), free-energetic driving forces, and variances in measured fluxes. 
Such measurements will enable inference-driven exploration of understudied and novel molecular machines and improve inference of the workhorse examples highlighted in this review.

Ultimately we should measure internal energy and information flows directly to test our theoretical predictions. Cutting-edge experiments are tantalizingly close to achieving the ability to track multiple degrees of freedom simultaneously with high spatiotemporal resolution. For example, dual-color MINFLUX shows significant promise~\cite{scheiderer2024dual}, allowing simultaneous tracking of both heads of a kinesin motor. As this data becomes available, theorists should develop and benchmark different methods for computing thermodynamic quantities from noisy, limited data. In the longer term, as measurement techniques improve and data becomes increasingly plentiful and precise, computational resources should be built for rapid, efficient, and tractable analysis.

\begin{tcolorbox}\textbf{FUTURE ISSUES}

\begin{enumerate}
\item Construct more detailed models of molecular machines to better match increasingly precise experimental data, and develop models to study other classes of molecular machines.
\item Analyze molecular machines in nontraditional circumstances, for example rapidly changing environments or active fluctuations.
\item Build theory to better understand multipartite molecular machines with complex interaction topologies.
\item Build intuition for the relationship between abstract information flows and concrete molecular motions, and explore the requirements to scale information engines beyond the nanoscale.
\item Experimentally measure key observables for understudied molecular machines, and derive more and better tools for thermodynamic inference.
\item Leverage emerging experimental techniques to simultaneously track multiple degrees of freedom with high spatiotemporal resolution, in order to directly measure internal energy and information flows.
\end{enumerate}
\end{tcolorbox}

%Disclosure
\section*{DISCLOSURE STATEMENT}
The authors are not aware of any affiliations, memberships, funding, or financial holdings that might be perceived as affecting the objectivity of this review.

% Acknowledgements
\section*{ACKNOWLEDGMENTS}
We thank Jordan Sawchuk, Johan du Buisson, Nancy Forde, John Bechhoefer (SFU Physics), Shoichi Toyabe (Tohoku Applied Physics), and Emanuele Penocchio (Northwestern Chemistry) for feedback on the manuscript. This work was supported by a Natural Sciences and Engineering Research Council of Canada (NSERC) CGS Doctoral fellowship (M.P.L.), an NSERC Discovery Grant and Discovery Accelerator Supplement RGPIN-2020-04950 (D.A.S.), and a Tier-II Canada Research Chair CRC-2020-00098 (D.A.S.).

% References
%
% Margin notes within bibliography
%\section*{LITERATURE\ CITED}

\bibliography{main.bib}

\end{document}